\documentclass{article}
\usepackage{amsmath}
\usepackage{graphicx}

\title{Scalar model of the glueball}
\author{Vladimir Dzhunushaliev
\thanks{
E-mail: dzhun@hotmail.kg}}
\date{}

\begin{document}
\maketitle

\begin{center}
\textit{
Dept. Phys. and Microel. Engineer., Kyrgyz-Russian
Slavic University\\
Bishkek, Kievskaya Str. 44, 720021, Kyrgyz
Republic}
\end{center}

\begin{abstract}
A scalar model of the glueball is offered. The model is based on the 
nonperturbative calculation of 2 and 4-points Green's functions. 
Approximately they can be expressed via a scalar field. On the basis 
of the SU(3) Yang-Mills Lagrangian an effective Lagrangian for the 
scalar field is derived. The corresponding field equations are solved 
for the spherically symmetric case. The obtained solution is interpreted 
as a bubble of the SU(3) quantized gauge field. 
\end{abstract}

\section{Introduction}

The nonlinearity of quantum chromodynamics leads probably to such 
objects as a hypothesized flux tube filled with a longitudinal 
color electric field and stretched between quark and antiquark and 
a glueball which is a blob of gluonic fields. 
There is a difference between flux tube and glueball: the first object 
is created by the quark and antiquark (the sources of the color field) 
the second one has not any sources - it is a blob of selfinteracting 
fields. The flux tube has the directed electro-color field but for the 
glueball it is not clear: whether is it something like monopole or no. 
These objects are impossible in linear theories, for instance in Maxwell 
electrodynamics, as its existence is connected with a nonlinear fields 
interaction among themselves. 
\par 
At the moment there are different glueball models such as flux tube 
\cite{isgur-paton}, bags \cite{donoghue-barnes}, constituent gluons 
\cite{barnes-coyne}, QCD Hamiltonian in Coulomb gauge \cite{szczepaniak}, 
non-Abelian Born-Infeld theory \cite{galtsov}, 
or the conjectured duality between supergravity 
and large-N gauge theories \cite{csaki-myers}. 
\par 
Here we present the glueball model where the gluon field is completely 
quantum one and it is described by the nonperturbative manner. In this 
model the gauge field $A^B_\mu$ is organized by such a way that 
$\left\langle A^B_\mu \right\rangle = 0$ but 
$\left\langle \left( A^B_\mu \right)^2 \right\rangle \neq 0$. In this case 
the color field in glueball performs nonlinear oscillations and the 
nonlinearity of the Yang-Mills equations do not allow us to present these 
oscillations as quanta, i.e. \textit{the glueball is not a cloud of quanta}. 
The basis for the presented glueball model is: (a) the initial SU(3) Lagrangian 
for the quantized field $\widehat{A}^B_\mu$ is averaged over some quantum 
state $\left| Q \right\rangle$; (b) the 2 and 4-points Green's functions 
$\biggl( \left\langle A^B_\mu A^C_\nu \right\rangle$ and 
$\left\langle A^B_\mu A^C_\nu A^D_\alpha A^E_\beta \right\rangle \biggl)$ 
arising in this case are approximated with help of some multiplet of scalar 
fields; (c) varying with respect to these scalar fields give rise to 
equations which describe these Green's functions. 

\section{$A^B_\mu \rightarrow \phi^B$ approximation}

In any quantum field theory the Green's functions give us the full 
information about quantized fields. In this section we would like to 
present equations which will describe 2 and 4-points Green's functions by 
some approximate manner in the QCD. For this we will average the SU(3) 
Lagrangian where we use some approximate expressions for 2 
and 4-points Green's functions. The SU(3) Lagrangian is 
\begin{equation}
  \widehat \mathcal{L}_{SU(3)} = \frac{1}{4} 
  \widehat F^A_{\mu \nu}\widehat F^{A \mu \nu}
\label{sec1:10}
\end{equation}
where $\widehat F^B_{\mu \nu} = \partial_\mu \widehat A^B_\nu - 
\partial_\nu \widehat A^B_\mu + g f^{BCD} \widehat A^C_\mu \widehat A^D_\nu$
is the field strength operator; $B,C,D = 1, \ldots ,8$ are the SU(3) color indices; 
$g$ is the coupling constant; $f^{BCD}$ are the structure constants for 
the SU(3) gauge group; $\widehat A^B_\mu$ is the gauge potential operator. 
In order to derive equations describing the quantized field we average 
the Lagrangian over a quantum state $\left.\left. \right| Q \right\rangle$ 
\begin{equation}
\begin{split}
	\left\langle Q \left| \widehat \mathcal{L}_{SU(3)}(x) \right| Q \right\rangle =
	\left\langle \widehat \mathcal{L}_{SU(3)} \right\rangle =& 
	\frac{1}{2}	
	\left\langle 
	  \left( \partial_\mu \widehat A^B_\nu (x) \right) 
	  \left( \partial^\mu \widehat A^{B\nu} (x) \right) - 
	  \left( \partial_\mu \widehat A^B_\nu (x) \right) 
	  \left( \partial^\nu \widehat A^{B\mu} (x) \right) 
	\right\rangle + \\
	&\frac{1}{2}	g f^{BCD} 
	\left\langle 
	  \left( \partial_\mu \widehat A^B_\nu (x)- 
	  \partial_\nu \widehat A^B_\mu (x)\right)
	  \widehat A^{C \mu} (x)\widehat A^{D \nu}(x) 
	\right\rangle + \\
	&\frac{1}{4}g^2 f^{BC_1D_1} f^{BC_2D_2}
	\left\langle 
	  \widehat A^{C_1}_\mu (x)\widehat A^{D_1}_\nu (x)
	  \widehat A^{C_2 \mu} (x)\widehat A^{D_2\nu} (x)
	\right\rangle .
\end{split}	
\label{sec1:20}
\end{equation}
One can see that schematically we have the following 2, 3 and 4-points 
Green's functions: 
$\left\langle \left( \partial A \right)^2 \right\rangle$, 
$\left\langle \left( \partial A \right) A^2 \right\rangle$ and 
$\left\langle \left( A \right)^4\right\rangle$. At first we suppose that 
the odd Green's functions can be written as the following product  
\begin{equation}
	\left\langle 
	  \widehat A^B_\alpha (x)\widehat A^C_\beta (y)\widehat A^D_\gamma (z)
	\right\rangle \approx 
	\left\langle 
	  \widehat A^B_\alpha (x)\widehat A^C_\beta (y)
	\right\rangle
	\left\langle 
	  \widehat A^D_\gamma (z)
	\right\rangle + \text{(other permutations)}
	= 0 
\label{sec1:30}
\end{equation}
as we have supposed that $\langle \widehat A^B_\alpha (x) \rangle = 0$. 
Later we suppose that 2-point Green's function can be presented in so called 
one-function approximation \cite{vdsin2} as 
\begin{equation}
	\left\langle 
	  \widehat A^B_\alpha (x) \widehat A^C_\beta (y) 
	\right\rangle = 
	\mathcal{G}^{BC}_{\alpha \beta} (x,y) 
  \approx 
	-\eta_{\alpha \beta} f^{BAD} f^{CAE} \phi^D (x) \phi^E(y)		
\label{sec1:35}	
\end{equation}
where $\phi^A(x)$ is the scalar field which describes the 2-point Green's 
function. The 4-point Green's function can be written in one-function 
approximation as the product of corresponding two 2-point Green's functions 
\begin{equation}
\begin{split}
	&\left\langle 
	  \widehat A^B_\alpha (x) \widehat A^C_\beta (y)
	  \widehat A^D_\gamma (z) \widehat A^R_\delta (u)
	\right\rangle \approx 
	\left\langle 
	  \widehat A^B_\alpha (x) \widehat A^C_\beta (y)
	\right\rangle  
	\left\langle 
	  \widehat A^D_\gamma (z) \widehat A^R_\delta (u)
	\right\rangle + \\
	&\left\langle 
	  \widehat A^B_\alpha (x) \widehat A^D_\gamma (z)
	\right\rangle  
	\left\langle 
	  \widehat A^C_\beta (y) \widehat A^R_\delta (u)
	\right\rangle + 	
	\left\langle 
	  \widehat A^B_\alpha (x) \widehat A^R_\delta (u)
	\right\rangle  
	\left\langle 
	  \widehat A^C_\beta (y) \widehat A^D_\gamma (z)
	\right\rangle .
\end{split}	
\label{sec1:110}
\end{equation}
Taking into account these expression for the 2,3 and 4-points Green's 
functions we can derive an effective Lagrangian 
$\mathcal {L}_{eff} = \left\langle \widehat \mathcal {L} \right\rangle$ 
for the scalar field $\phi^A$ which describes 2 and 4-points Green's 
functions (for details, see Appendix \ref{app1})
\begin{equation}
\begin{split}
	&\mathcal {L}_{eff} = - \frac{1}{2}\left( \partial_\mu \phi^A \right)^2 + 
	\frac{\lambda_1}{4} 
	\left[ \phi^a \phi^a - \phi^a_0 \phi^a_0 
	\right]^2 - \frac{\lambda_1}{4} \left( \phi^a_0 \phi^a_0  \right)^2 + \\
	&\frac{\lambda_2}{4} 
	\left[ \phi^m \phi^m - \phi^m_0 \phi^m_0 
	\right]^2 - \frac{\lambda_2}{4} \left( \phi^m_0 \phi^m_0  \right)^2 + 
	\left( \phi^a \phi^a \right) \left( \phi^m \phi^m \right) 
\end{split}
\label{sec1:260}
\end{equation}
where the indices $a=1,2,3$ are $SU(2)$ indices, $m=4,5,6,7,8$ are 
the coset $SU(3)/SU(2)$ indices, $\phi^A_0$ are some constants. 
The field equations are 
\begin{eqnarray}
  \partial_\mu \partial^\mu \phi^a &=& 
  -\phi^a \left[ 2 \phi^m \phi^m + \lambda_1 
  \left(
    \phi^a \phi^a- \phi^a_0 \phi^a_0 
  \right) \right],
\label{sec1:270}\\
  \partial_\mu \partial^\mu \phi^m &=& 
  -\phi^m \left[ 2 \phi^a \phi^a + \lambda_2 
  \left(
    \phi^m \phi^m- \phi^m_0 \phi^m_0 
  \right) \right].
\label{sec1:280}
\end{eqnarray}
One can note that in Ref. \cite{bazeia} was investigated a system 
of coupled scalar fields and was shown that such system may have 
soliton solutions by some specific choice of the potential term. 

\section{Glueball as a bubble of $\phi^A$ field}

Now we would like to consider the spherically symmetric solution with 
the following ansatz for the scalar field 
\begin{eqnarray}
	\phi^a (r) &=& \frac{\phi(r)}{\sqrt{6}} , \quad a=1,2,3 ,
\label{sec2:10}\\
  \phi^m (r) &=& \frac{f(r)}{\sqrt{10}} , \quad m=4,5,6,7,8 .
\label{sec2:20}
\end{eqnarray}
Let us note that this ansatz means that the components $\phi^a$ have 
another behaviour then the components $\phi^m$. 
One can say that such situation is close to a colored flux tube 
\cite{dzhun1} solution filled with the longitudinal electric field. 
After substitution \eqref{sec2:10} \eqref{sec2:20} into equation 
\eqref{sec1:270} \eqref{sec1:280} we have 
\begin{eqnarray}
	\phi'' + \frac{2}{r} \phi' &=& \phi 
	\left[ 
	  f^2 + \lambda_1 \left( \phi^2 - m^2 \right)
	\right], 
\label{sec2:30}\\
	f'' + \frac{2}{r} f' &=& f 
	\left[ 
	  \phi^2 + \lambda_2 \left( f^2 - \mu^2 \right)
	\right] 
\label{sec2:40}  
\end{eqnarray}
where $2\phi_0^a \phi_0^a = m^2$ and 
$2 \phi_0^m \phi_0^m = \mu^2$; 
$m, \mu$ are some constants which will be calculated by solving  
equations \eqref{sec2:30} and \eqref{sec2:40} and we redefine 
$\lambda_{1,2} / 2 \rightarrow \lambda_{1,2}$. 
Evidently these equations can not be calculated analytically. The 
preliminary numerical investigations show that this equations set 
do not have regular solutions by arbitrary choice of $m, \mu$ parameters. 
We will solve equations \eqref{sec1:270} \eqref{sec1:280} as a nonlinear 
eigenvalue problem for eigenstates $\phi(x), f(x)$ and eigenvalues 
$m, \mu$, i.e. we will calculate $m, \mu$ parameters such that 
the regular functions $\phi(r)$ and $f(r)$ do exist. 
\par 
At first we note that the solution depends on the following parameters: 
$\phi(0), f(0)$ and $\lambda_{1,2}$. We can decrease the number of these 
parameters dividing equations \eqref{sec2:30} \eqref{sec2:40} on 
$\phi^3(0)$. After this we introduce the dimensionless radius 
$x=r\phi(0)$ and redefine 
$\phi(x)/\phi(0) \rightarrow \phi(x)$, 
$f(x)/\phi(0) \rightarrow f(x)$ and $m/\phi(0) \rightarrow m$, 
$\mu/\phi(0) \rightarrow \mu$. Thus we have the following equations set 
\begin{eqnarray}
	\phi'' + \frac{2}{x} \phi' &=& \phi 
	\left[ 
	  f^2 + \lambda_1 \left( \phi^2 - m^2 \right)
	\right], 
\label{sec2:50}\\
	f'' + \frac{2}{x} f' &=& f 
	\left[ 
	  \phi^2 + \lambda_2 \left( f^2 - \mu^2 \right)
	\right]. 
\label{sec2:60}  
\end{eqnarray}
We will search the regular solution with the following boundary 
conditions 
\begin{eqnarray}
	\phi(0) &=& 1, \quad \phi(\infty) = m ,
\label{sec2:70}\\
  f(0) &=& f_0, \quad f(\infty) = 0 .
\label{sec2:80}
\end{eqnarray}
One can say that $\phi(x)$ is like to kink and $f(x)$ to soliton. 
Let us rewrite equation \eqref{sec2:60} in the following form 
\begin{equation}
	-\left( f'' + \frac{2}{x} f' \right) + f V_{eff} = 
	\left( \lambda_2 \mu^2 \right) f
\label{sec2:90}
\end{equation}
where we have introduced an effective potential 
\begin{equation}
	V_{eff} = \left( \phi^2 + \lambda_2 f^2 \right).
\label{sec2:100}
\end{equation}
Immediately we see that with the boundary conditions \eqref{sec2:80} 
equation \eqref{sec2:90} is the Schr\"odinger equation and it 
may have a regular solution only if $V_{eff}$ has a hole 
$(\phi^2 \stackrel{x\rightarrow \infty}{\longrightarrow} const , 
f^2 \stackrel{x\rightarrow \infty}{\longrightarrow} 0)$ and an 
energy level $\lambda_2 \mu^2$  have to be quantized. 

\section{Numerical solution}

We choose the following numerical method for solving equations 
\eqref{sec2:50} \eqref{sec2:60}: we take a null approximation for 
the function $f(x)$ (which is $f_0(x)$) and solve equation \eqref{sec2:50} 
in the followng form
\begin{equation}
	\phi_0'' + \frac{2}{x} \phi_0 ' = \phi_0 
	\left[ 
	  f^2_0 + \lambda_1 \left( \phi^2_0 - m^2_0 \right)
	\right]
\label{sec3:10}
\end{equation}
where $m_0$ is the null approxiamtion for the parameter $m$, the 
boundary conditions are \eqref{sec2:70} and the function 
$\phi_0(x)$ is zero approximation for the function $\phi(x)$. 
Having the regular solution $\phi_0(x)$ we can
substitute it into equation \eqref{sec2:60} and solve the 
equation
\begin{equation}
	f_1'' + \frac{2}{x} f_1' = f_1 
	\left[ 
	  \phi^2_0 + \lambda_2 \left( f^2_1 - \mu^2_1 \right)
	\right]
\label{sec3:20}
\end{equation}
with the boundary conditions \eqref{sec2:80} (for the numerical calculations 
presented here we take $f_0 = \sqrt{0.6}$). Thus we have the first 
approximation $f_1(x)$ which we substitute into equation \eqref{sec2:50} 
\begin{equation}
	\phi_1'' + \frac{2}{x} \phi_1 ' = \phi_1 
	\left[ 
	  f^2_1 + \lambda_1 \left( \phi^2_1 - m^2_1 \right)
	\right].
\label{sec3:30}
\end{equation}
This equation gives us the first approximation for the function 
$\phi_1(x)$ and so on. On the $i^{th}$ step we will have 
\begin{equation}
	\phi_i'' + \frac{2}{x} \phi_i ' = \phi_i 
	\left[ 
	  f^2_{i-1} + \lambda_1 \left( \phi^2_i - m^2_i \right)
	\right]
\label{sec3:40}
\end{equation}
and 
\begin{equation}
	f_i'' + \frac{2}{x} f_i' = f_i 
	\left[ 
	  \phi^2_i + \lambda_2 \left( f^2_i - \mu^2_i \right)
	\right].
\label{sec3:50}
\end{equation}
For every step we have the values $m^2_i$ and $\mu^2_i$ as an 
approximation for the true eigenvalues values ${m^*}^2$ and ${\mu^*}^2$. 

\subsection{The more detailed description of the numerical calculations}

\begin{figure}[h]
  \begin{minipage}[t]{.45\linewidth}
  \begin{center}
    \fbox{
    \includegraphics[height=5cm,width=5cm]{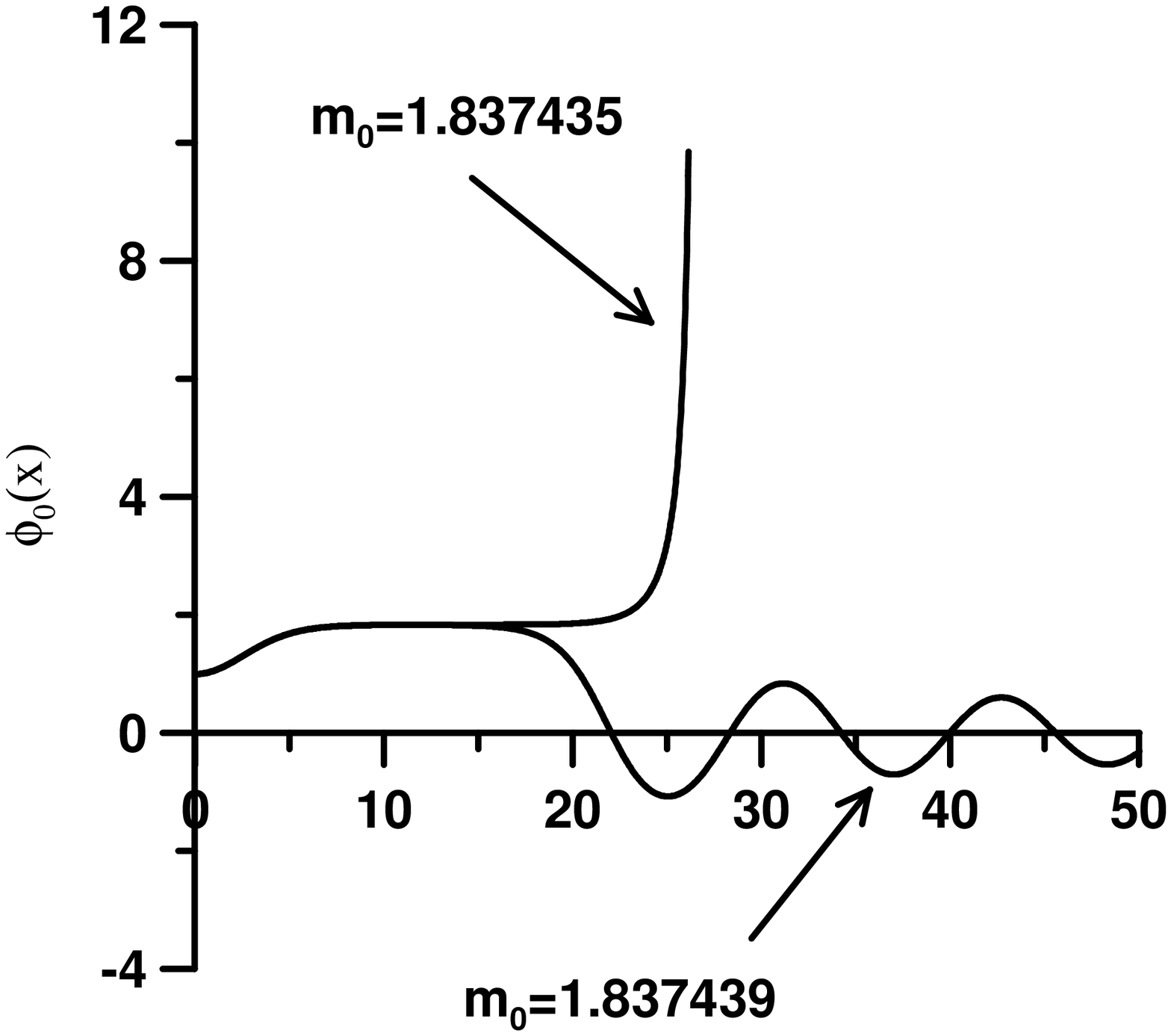}}
    \caption{The typical singular solutions for equation \eqref{sec3:10}.
    $\lambda_1=0.1$.}
    \label{fig:phi-sing}
  \end{center}  
  \end{minipage}\hfill
  \begin{minipage}[t]{.45\linewidth}
  \begin{center}
    \fbox{
    \includegraphics[height=5cm,width=5cm]{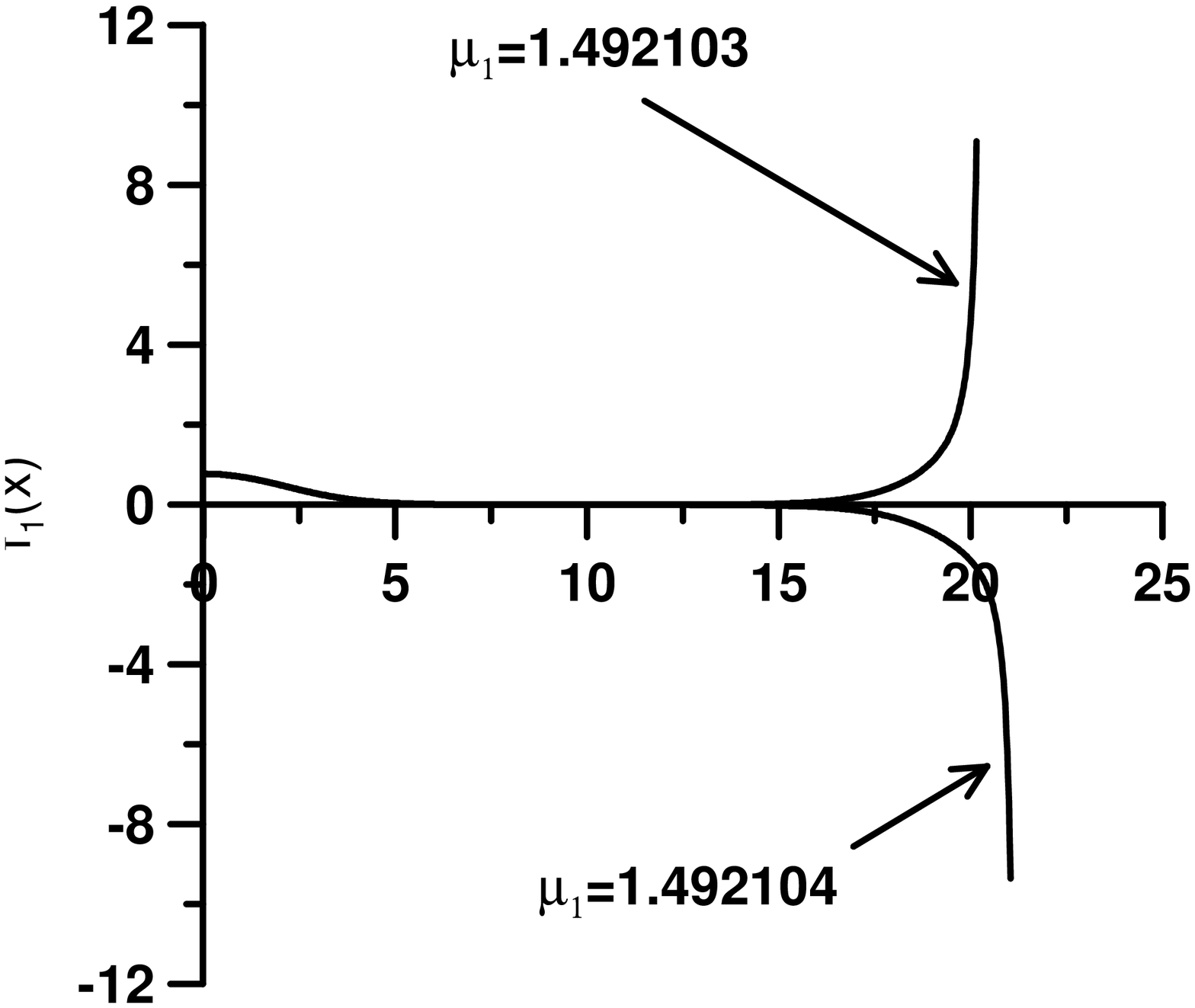}}
    \caption{The typical singular solutions for equation \eqref{sec3:20}.
    $\lambda_2=1.0$.}
    \label{fig:f-sing}
  \end{center}  
  \end{minipage} 
\end{figure}

At first we will describe the numerical solution of equation \eqref{sec3:10}. 
For this we choose the null approximation for $f(x)$ as 
\begin{equation}
	f_0(x) = \frac{\sqrt{0.6}}{\cosh^2{\frac{x}{4}}}.
\label{sec3a:10}
\end{equation}
The typical solutions for the arbitrary value of $m_0$ are presented on 
Fig.\ref{fig:phi-sing}. 
We see that by $m_0 < m_0^*$ ({$m_0^*$ is an unknown parameter 
which gives us the regular solution) the solution $\phi_0(x)$ is 
singular and near to the singularity the equation has the form 
\begin{equation}
	\phi_0'' \approx \lambda_1 \phi_0^3
\label{sec3a:20}
\end{equation}
consequently the solution is 
\begin{equation}
	\phi_0(x) \approx \sqrt{\frac{2}{\lambda_1}} \frac{1}{x_0 - x}
\label{sec3a:30}
\end{equation}
where $x_0$ is some constant depending on $m_0$. On the other hand 
by $m_0 > m_0^*$ the solution is presented on Fig.\ref{fig:phi-sing}  
and the corresponding asymptotical equation is 
\begin{equation}
	\phi_0''(x) + \frac{2}{x}	\phi_0' \approx - 
	\left( \lambda_1 m^2 \right) \phi_0
\label{sec3a:40}
\end{equation}
which has the following solution 
\begin{equation}
	\phi_0(x) \approx \phi_\infty 
	\frac{\sin{\left(x \sqrt{\lambda_1 m^2} + \alpha\right)}}{x}
\label{sec3a:50}
\end{equation}
where $\phi_\infty$ and $\alpha$ are some constants. All of that allows us to 
assert that there is a value $m^*_0$ for which does exist an exceptional solution 
which with some accuracy is presented on Fig.\ref{fig:phi-reg}. For this value 
$m^*_0$ the equation \eqref{sec3:10} has the following asymptotical 
behaviour 
\begin{equation}
	\phi_0''(x) + \frac{2}{x}	\phi_0' \approx 
	2 \lambda_1 \left( m^*_0 \right)^2 
	\left( \phi_0 - m^*_0 \right)
\label{sec3a:60}
\end{equation}
and the corresponding asymptotical solution is 
\begin{equation}
	\phi_0(x) \approx m^*_0 + \beta_0 
	\frac{e^{-x \sqrt{2 
	\lambda_1 \left( m^*_0 \right)^2}}}{x}
\label{sec3a:70}
\end{equation}
where $\phi_\infty$ and $\beta$ are some constants. 
\par 
The next step is finding the first approximation for the $f_1(x)$ 
function. The equation is 
\begin{equation}
  f_1'' + \frac{2}{x} f' = f_1 
  \left[  
    \phi_0^2 + \lambda_2 \left( f_1^2 - \mu^2_1 \right)
  \right].
\label{sec3a:80}
\end{equation}
From the previous calculations one can assume that there is the exceptional 
regular 
solution $\phi_0(x)$ with the asymptotical behaviour \eqref{sec3a:70}. 
Then the numerical investigation shows that for the arbitrary $\mu$ 
there are two different singular solutions which are presented 
on Fig.\ref{fig:f-sing}. 
\begin{figure}[h]
  \begin{minipage}[t]{.45\linewidth}
  \begin{center}
    \fbox{
    \includegraphics[height=5cm,width=5cm]{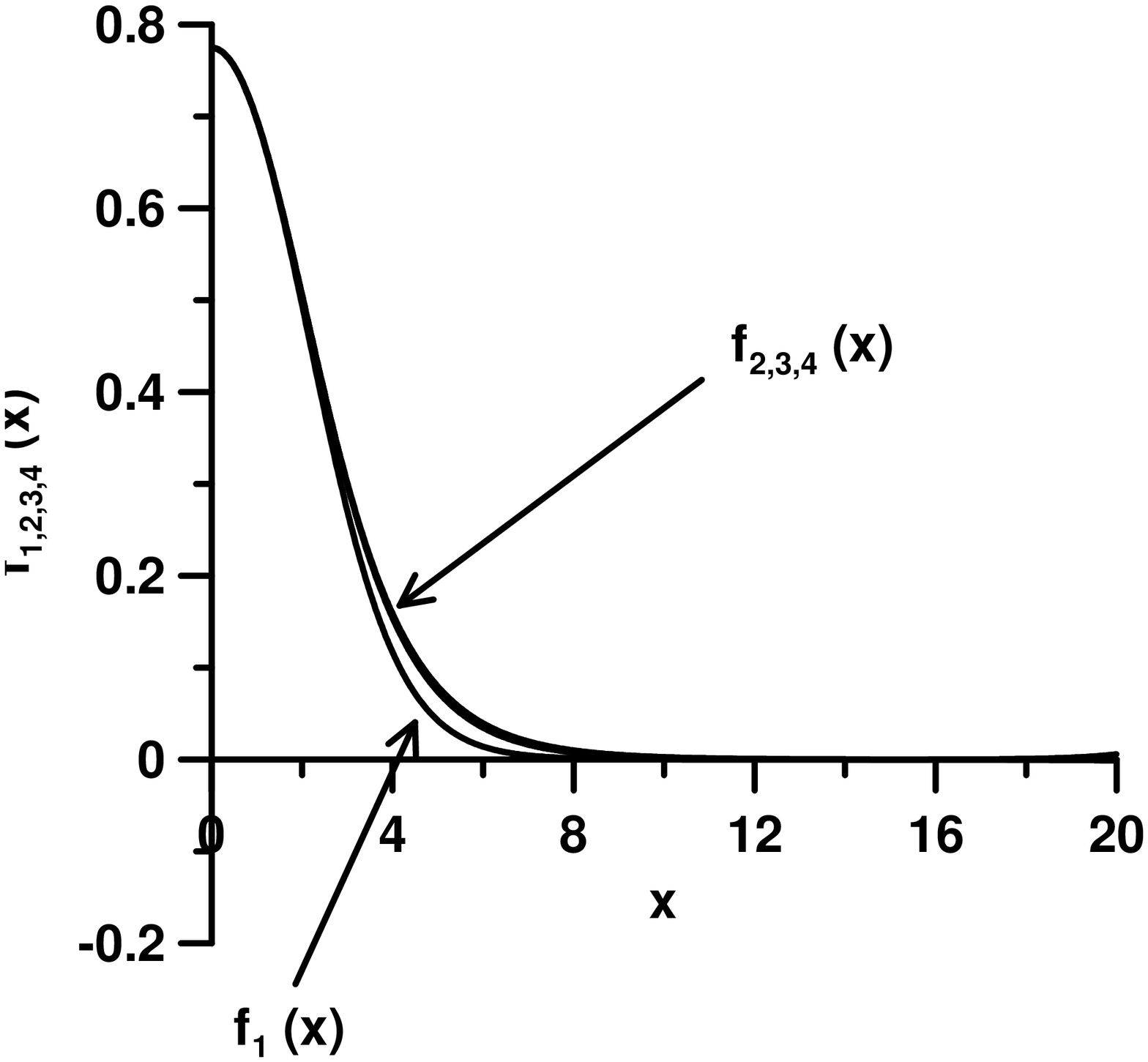}}
    \caption{The iterative functions $f_{1,2,3,4}(x)$.}
    \label{fig:f-reg}
  \end{center}  
  \end{minipage}\hfill
  \begin{minipage}[t]{.45\linewidth}
  \begin{center}
    \fbox{
    \includegraphics[height=5cm,width=5cm]{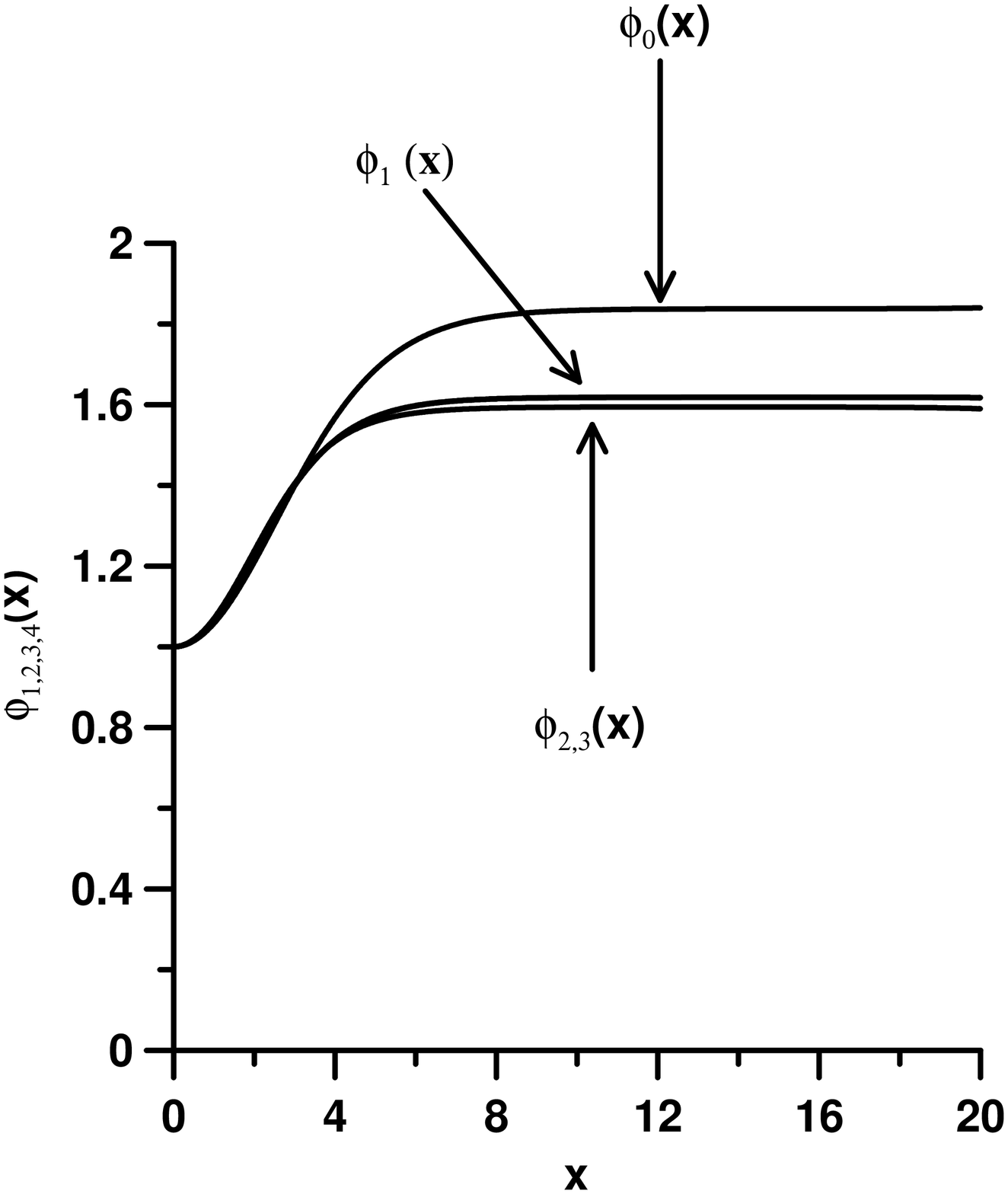}}
    \caption{The iterative functions $\phi_{1,2,3,4}(x)$.}
    \label{fig:phi-reg}
  \end{center}  
  \end{minipage} 
\end{figure}
Analogously to equation \eqref{sec3a:20} the singular 
behaviour of the function $f_1(x)$ is 
\begin{eqnarray}
	f_1(x) &\approx & \sqrt{\frac{2}{\lambda_2}} \frac{1}{x - x_0} 
	\quad \text {by} \quad \mu_1 < \mu^*_1 ,
\label{sec3a:90}\\
  f_1(x) &\approx & - \sqrt{\frac{2}{\lambda_2}} \frac{1}{x - x_0} 
	\quad \text {by} \quad \mu_1 > \mu^*_1 .
\label{sec3a:100}
\end{eqnarray}
Evidently that we can suppose that there is a regular exceptional 
solution $f^*_1(x)$ by $\mu_1 = \mu^*_1$ with the following asymptotical 
behaviour 
\begin{equation}
	f^*_1(x) \approx f_{\infty ,1}
	\frac{e^{- x \sqrt{\left( m_0^* \right)^2 - \lambda_2 \left( \mu_1^* \right)^2}}}{x} 
\label{sec3a:110}
\end{equation}
where $f_\infty$ is some parameter. 
The next step is substituting the first approximation $f^*_1(x)$ into 
equation \eqref{sec2:50} for finding the regular exceptional 
solution $\phi^*_1(x)$ by $m^*_1$ then $\phi^*_1(x)$ will be substituted 
into equation \eqref{sec2:60} for finding the regular exceptional 
solution $f^*_2(x)$ by $\mu = \mu^*_2$ and so on. 
\par 
The result of these calculations is presented on 
Fig's.\ref{fig:f-reg}, \ref{fig:phi-reg} and 
Table \ref{table1}. We see that there is the convergence 
$\phi^*_i(x) \rightarrow \phi^*(x)$, 
$f^*_i(x) \rightarrow f^*(x)$, $m^*_i \rightarrow m^*$ and 
$\mu^*_i \rightarrow \mu^*$ where $f^*(x), \phi^*(x)$ are the eigenstates 
and $m^*, \mu^*$ are eigenvalues of nonlinear eigenvalue problem 
\eqref{sec2:30} \eqref{sec2:40}. 
\par
The verification of the presented numerical method was done for the 
soliton solution, for details see Appendix \ref{app2}.
\begin{table}[h]
    \begin{center}
        \begin{tabular}{|c|c|c|c|c|}\hline
          i & 1 & 2& 3 & 4 \\ \hline
            $m^*_i$ & 1.8374351\ldots & 1.594328\ldots & 1.6186108\ldots 
            & 1.61823766\ldots \\ \hline
            $\mu^*_i$ & 1.492105312\ldots & 1.4938287\ldots & 1.4921473\ldots 
            & 1.4921473\ldots \\ \hline
        \end{tabular}
    \end{center}
    \caption{The iterative parameters $m^*_i$ and $\mu^*_i$.}
    \label{table1}
\end{table}

\section{The properties of solution}

In this section we would like to describe the properties of the derived solution. 
It is easy to see that the asymptotical behaviour of the regular solution is 
\begin{eqnarray}
	\phi^*(x) \approx m^* + \beta	\frac{e^{-x \sqrt{2 
	\lambda_1 \left( m^* \right)^2}}}{x}
\label{sec5:10}\\
	f^*(x) \approx f_\infty
	\frac{e^{- x \sqrt{\left( m^* \right)^2 - \lambda_2 \left( \mu^* \right)^2}}}{x} 
\label{sec5:20}
\end{eqnarray}
where $m^*$ and $\mu^*$ are the parameters derived in the coarse of iterative solution 
of equations \eqref{sec2:50} \eqref{sec2:60}. 
The energy density  of the presented solution is 
\begin{equation}
	\varepsilon(r) = \frac{1}{g^2}
	\left[
		{\phi'}^2(r) + {f'}^2(r) + \frac{\lambda_1}{2} \left( \phi^2(r) - {m^*}^2 \right) + 
		\frac{\lambda_2}{2} f^2(r) \left( f^2(r) - 2 {\mu^*}^2 \right) + 
		f^2(r) \phi^2(r)	  	
	\right]
\label{sec5:25}
\end{equation}
here $\lambda_{1,2}, m^*$ and $\mu^*$ are redefined according the remark after eq. 
\eqref{sec2:40} and we add the constant term $-\lambda_2 \mu^2 /4$ for the finiteness 
of the full energy. Thus the glueball energy is 
\begin{equation}
	W = \frac{4 \pi}{g^2} \phi_0 \int\limits^\infty_0 x^2 
	\left[
	  {f'}^2 + {\phi'}^2 + \frac{\lambda_1}{2} \left( \phi^2 - {m^*}^2 \right)^2 + 
	  \frac{\lambda_2}{2} f^2 \left( f^2 - 2{\mu^*}^2 \right) + 
	  f^2 \phi^2
	\right] dx = 
	\frac{4 \pi}{g^2} \phi_0 I_1\left( \lambda_{1,2}, m^*, \mu^* \right)
\label{sec5-30}
\end{equation}
here we have redefined $r, f$ and $\phi$ according to remark before eq. \eqref{sec2:50}. 
The quantity $\phi_0^{-1}$ defines the radius of flux tube since 
the dimensionless variables $x$ for flux tube is $x = \rho \phi_0$. 
The profile of the energy density is presented on Fig. \ref{fig:energy}. 
The numerical calculations for the dimensionless integral $I_1$ gives 
\begin{equation}
	I_1 = \int\limits^\infty_0 x^2 
	\left[
	  {f'}^2 + {\phi'}^2 + \frac{\lambda_1}{2} \left( \phi^2 - {m^*}^2 \right)^2 + 
	  \frac{\lambda_2}{2} f^2 \left( f^2 - 2{\mu^*}^2 \right) + 
	  f^2 \phi^2
	\right] dx \approx 6.28 .
\label{sec5:40}
\end{equation}
\begin{figure}[h]
  \begin{center}
    \fbox{
    \includegraphics[height=5cm,width=5cm]{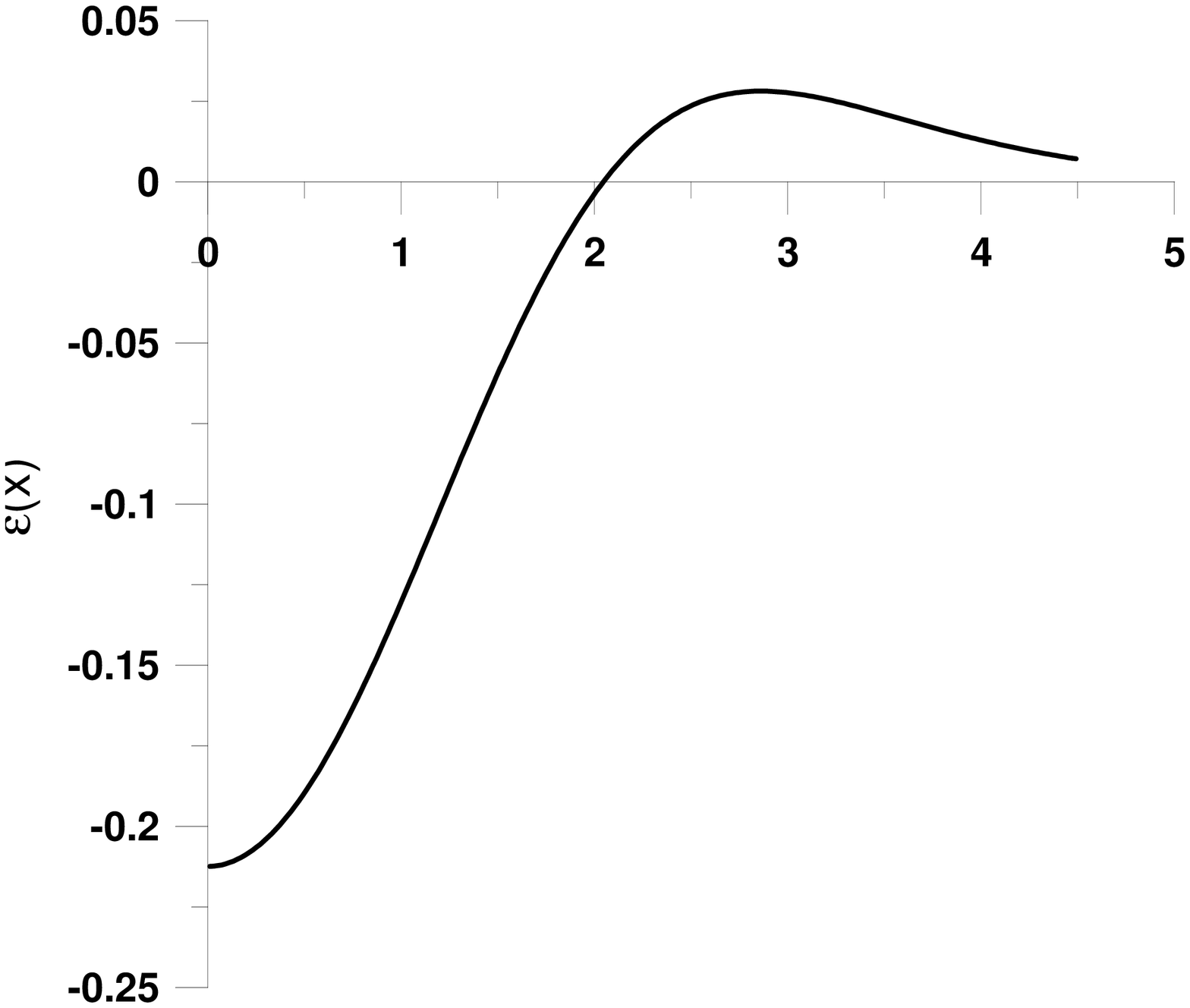}}
    \caption{The profile of the energy density.}
    \label{fig:energy}
  \end{center}  
\end{figure}
\par 
The next what we can do is the calculation the ratio of the energy glueball 
and string tension. On the basis of similar ideas presented here a flux tube 
solution was found in Ref. \cite{dzhun1}. The basic idea is that the 
quantized SU(3) gauge potential $A^B_\mu, B=1,2, \ldots, 8$ can be 
splitted on two pieces: (a) the potential components $A^a_\mu, a=1,2,3$ 
belong to the SU(2) subgroup, (b) the potential components 
$A^m_\mu, m=4,5,6,7,8$ are coset componets. For the flux tube case it is supposed 
that $A^a_\mu$ are almost classical degrees of freedom but $A^m_\mu$ componets 
(analogously to glueball case) are completely quantum degrees of fredom. 
2 and 4-points Green's functions for the $A^m_\mu$ components are expressed via 
a scalar field $\phi^a$ 
(what is similar to presented here glueball case). The assumptions which are 
similar to glueball case lead to a numerical solution describing a flux tube 
filled with the longitudinal color electric field. The linear energy density 
(or string tension) is 
\begin{equation}
\begin{split}
	\sigma = &\frac{\pi}{g^2} \int\limits^\infty_0 \rho 
	\left[
	  {f'}^2 + {v'}^2 + {\phi'}^2 + v^2 f^2 + v^2\phi^2 + f^2 \phi^2 + 
	  m_1^* f^2 - m_2^*v^2 + \frac{\lambda}{2} 
	  \left(
	    \phi^2 - \phi^*_\infty
	  \right)^2
	\right] d\rho = \\ 
	& \frac{\pi}{g^2} \phi_0^2 I_2\left( \lambda, \phi_\infty, m^*_{1,2} \right)
\end{split}
\label{sec5:90}
\end{equation}
where 
\begin{equation}
    A^1_t(\rho) = \frac{f(\rho)}{g} ; \quad A^2_z(\rho) = \frac{v(\rho)}{g} ;
    \quad \phi^3(\rho) = \frac{\phi(\rho)}{g}.
\label{sec5:100}
\end{equation}
The numerical calculations give $I_2 \approx 0.63$. 
\par
The flux of the electric field is 
\begin{equation}
    \Phi = \int E^3_z ds = 2 \pi \int\limits^\infty_{0} \rho \frac{f(\rho)v(\rho)}{g}
    d\rho = \frac{2 \pi}{g} \int\limits^\infty_0 x f(x) v(x) dx = 
    \frac{2 \pi}{g} I_3 \left( \lambda, \phi_\infty , m^*_{1,2} \right)
\label{sec5:110}
\end{equation}
where $E^3_z = fv/g$ is the longitudinal color electric field. Eq. \eqref{sec5:110} 
shows that (like to Coulomb law) the flux of the electric field is proportional to 
a color charge defined as $q = 1/g$. But of course there is a dimensionless 
correction $I_3$ coming from the nonlinearity of the theory. 
The numerical calculations give $I_3 \approx 0.79$. 
\par
Let us consider the ratio
\begin{equation}
	\frac{W}{\sqrt{\sigma}} = \sqrt{\frac{4 \pi}{g^2}} \; \frac{2I_1}{\sqrt{I_2}} 
	\approx 5
\label{sec5:120}
\end{equation}
here we take into account the value of dimensionless constant $g^2/4\pi \approx 10$. 
It is necessary to note that for the ratio \eqref{sec5:120} we consider the case when the scalar 
field $\phi$ has the same magnitudes at the center of flux tube and glueball. It can be 
compared with the lattice calculations \cite{teper} where this quantity is presented 
as $W/\sqrt{\sigma} \approx 3.64$ for $0^{++}$ glueball. 
\par
It is interesting to consider the dimensionless ratio 
\begin{equation}
	\frac{\sqrt{\sigma} \Phi}{W} = \frac{\sqrt{\pi}}{4} \frac{\sqrt{I_2} I_3}{I_1} 
	\approx 0.04
\label{sec5:130}
\end{equation}
which tell us that the ratio \eqref{sec5:120} is proportional to the flux of 
electric field. 
\par 
Now we would like to calculate the angular momentum of the glueball in the offered 
model. The angular momentum operator is 
\begin{equation}
	\widehat{\vec{M}} = \int \left[ \vec r \times 
	\left[ \hat{\vec E}^A \times \hat{\vec H}^A \right] \right] 
	dV
\label{sec5:60}
\end{equation}
where $\hat{E}^B_i = \hat{F}^B_{0i} = \partial_0 \hat{A}^B_i - \partial_i \hat{A}^B_0 + 
g f^{BCD} \hat{A}^C_0 \hat{A}^D_i$ is the the operator of the color electric field; 
$\hat{H}^B_i = \epsilon_{ijk}\hat{F}^{Bjk}$ is the operator of the color magnetic field, and  
$\hat{F}^B_{jk} = \partial_j \hat{A}^B_k - \partial_k \hat{A}^B_j + 
g f^{BCD} \hat{A}^C_j \hat{A}^D_k$; $i,j,k = 1,2,3$. 
Let us consider 
\begin{equation}
\begin{split}
	&\hat{m}_i = \left[ \vec r \times \left[ \hat{\vec E}^A \times \hat{\vec H}^A \right] \right]_i = 
	\epsilon_{ijk} \epsilon_{klm} \epsilon_{mpq} x^j 
	\hat{F}^A_{0l} \hat{F}^{Apq} = \\
	&\epsilon_{ijk} \epsilon_{klm} \epsilon_{mpq} x^j 
	\left(
	  \partial_0 \hat{A}^B_i - \partial_i \hat{A}^B_0 + g f^{BCD} \hat{A}^C_0 \hat{A}^D_i
	\right)
	\left(
	  \partial_j \hat{A}^B_k - \partial_k \hat{A}^B_j + g f^{BCD} \hat{A}^C_j \hat{A}^D_k
	\right)
\end{split}
\label{sec5:70}
\end{equation}
One can show that the expectation value 
\begin{equation}
	\left\langle \hat{m}_i \right\rangle = 0
\label{sec5:80}
\end{equation}
as in the considered case either $\partial_0 (\cdots) \equiv 0$ or 
$\left\langle \hat{A}^B_0 \hat{A}^C_i \cdots \right\rangle \equiv 0$ 
in the consequence of the presence the factor $\eta_{\mu \nu}$ 
in the assumptions \eqref{sec1:35} and \eqref{sec1:110}. 
It means that the spin of the presented glueball model is zero. 
It is interesting to note that there is an opinion \cite{single} that 
pure glueball can only be spin 0. 
\par 
From this consideration immediately we see that in this approach the glueball  
with nonzero spin probably can be derived using ans\"atz similar \eqref{sec1:35} 
but with nonzero correlation between $A^B_0$ and $A^C_i$. 
\par
Another interesting 
possibility for the future investigations is the derivation of a mass spectrum. 
For this we see two ways: the first one is the search of excited states on the 
basis of ans\"atz \eqref{sec1:35} for the 2-point Green's function; the second one 
is the search an another ans\"atz for the 2-point Green's function which gives 
the glueball with another mass. The preliminary investigations show that probably 
excited states can not be derived using the presented iterative method. Thus the  
derivation of glueball mass spectrum in the nonperturbative approach is the 
complicated problem and it is the goal of the future investigations. 
\par 
Finally, we would like to touch upon the connection between our solution 
and Derrick's Theorem \cite{derrick}. This theorem tells us that in 3 spatial 
dimensions the scalar field theory with nonnegative potential 
do not have absolute stable solutions with finite energy which means that at the 
infinity the solution must tend to a global minimum where $V(global\; minimum)=0$. 
But the potential for the
interacting scalar fields of \eqref{sec1:260} has global and
local minima. Our solution is in one of the local minima. The Derrick's Theorem 
tells us that if we add two constant terms $\frac{\lambda_1}{4} (\phi^a_0 \phi^a_0)^2$ 
and $\frac{\lambda_2}{4} (\phi^m_0 \phi^m_0)^2$ to the potential in 
eq. \eqref{sec1:260} we will have the following potential 
\begin{equation}
\begin{split}
	&V = \frac{\lambda_1}{4} 
	\left[ \phi^a \phi^a - \phi^a_0 \phi^a_0 
	\right]^2 + 
	\frac{\lambda_2}{4} 
	\left[ \phi^m \phi^m - \phi^m_0 \phi^m_0 
	\right]^2 + 
	\left( \phi^a \phi^a \right) \left( \phi^m \phi^m \right) , \\
	&V\left( \text{global minimum} \right) = 0 , \\
	&V\left( \text{local minimum} \right) > 0 .
\label{sec5:140}
\end{split}
\end{equation}
In this case the presented solution (in the full agreement with the Derrick's Theorem) 
will have the infinite energy. If we add only one constant term 
$\frac{\lambda_2}{4} (\phi^m_0 \phi^m_0)^2$ we will have the following potential 
\begin{equation}
\begin{split}
	&V = \frac{\lambda_1}{4} 
	\left[ \phi^a \phi^a - \phi^a_0 \phi^a_0 
	\right]^2 + 
	\frac{\lambda_2}{4} \phi^m \phi^m 
	\left[ \phi^m \phi^m - 2 \phi^m_0 \phi^m_0 
	\right] + 
	\left( \phi^a \phi^a \right) \left( \phi^m \phi^m \right) , \\
	&V\left( \text{global minimum} \right) < 0 , \\
	&V\left( \text{local minimum} \right) = 0 .
\label{sec5:150}
\end{split}
\end{equation}
In this case the energy of the glueball solution is finite but the stability of the 
solution have to be investigated in the future works. 
\par 
After the discussion of the properties of derived solution one can see that 
the presented model is the model of glueball with zero spin where some combination 
of fields $\phi^a$ and $\phi_m$ push out each other. 

\section{Physical discussion and conclusions}

In this letter we discuss the glueball solution presented on 
Fig's.\ref{fig:f-reg}, \ref{fig:phi-reg}. 
This scalar model of glueball is derived with the assumptions that: 
(a) 2 and 4-points Green's functions of the SU(3) gauge potential 
can be approximately expressed via a scalar field; (b) the scalar fields 
components with a small subgroup indices belonging to $SU(2) \in SU(3)$ 
can have the different 
qualitative behaviour in some physical situations (in flux tube and glueball) 
in comparison with the scalar components which indices belong 
to the coset $SU(3)/SU(2)$. As the consequence we see that in this case exists 
a blob of the quantized SU(3) gauge field and the coset components push out 
the SU(2) components of the scalar field that is like to the Meissner 
effect in superconductivity. Such solution can be interpreted as the glueball 
in a medium. It follows from the fact that 2 and 4-points 
Green's functions of $\phi^m$ are nonzero at the infinity. 
\par 
Remarkably that similar situation exists in a flux tube solution obtained 
in Ref.\cite{dzhun1}. There is only one essential difference between 
flux tube and glueball solutions: in the flux tube solution the gauge 
potential components belonging to the small subgroup SU(2) in the 
first approximation can be considered as classical degrees of freedom 
that allows to exist a longitudinal color electric field directed from 
quark to antiquark. 
\par 
Let us to underscore that in this interpretation the derived bubble of 
the quantized SU(2) components live in the sea of the quantized coset 
components. But we can present an another interpretation of this solution 
in which both components exist in vacuum.
\par 
One can unite two assumptions about 2-point Green's function 
\eqref{sec1:35} and the term breaking the gauge invariance 
$\phi_0^A \phi_0^A$ in one
\begin{equation}
	\left\langle 
	  \widehat A^B_\alpha (x) \widehat A^C_\beta (y) 
	\right\rangle 
  \approx 
	-\eta_{\alpha \beta} 
	\left[ 
	  f^{BAd} f^{CAe} \phi^d (x) \phi^e(y)	+ 
		f^{BAm} f^{CAn} 
			\left( \phi^m (x) \phi^n(y) - \phi^m_0 \phi^n_0
			\right)
	\right].
\label{sec4:10}	
\end{equation}
Let us remind that the indices $d,e = 1,2,3$ and $m,n =4,5,6,7,8$. 
In this case 
\begin{eqnarray}
	\left\langle 
	  \widehat A^a_\alpha (x) \widehat A^a_\beta (x) 
	\right\rangle 
  & \approx & 
	-\eta_{\alpha \beta} 
	\left[ 
	  \sum_{c}\left( f^{acd} \right)^2 \left( \phi^d (x) \right)^2	+ 
		\sum_{m}\left( f^{amn} \right)^2 
			\left( \phi^n (x) - \phi^n_0 
			\right)^2
	\right],
\label{sec4:20}\\	
  \left\langle 
	  \widehat A^m_\alpha (x) \widehat A^m_\beta (x) 
	\right\rangle 
  & \approx & 
	-\eta_{\alpha \beta} 
	\left[ 
	  \sum_{n}\left( f^{mna} \right)^2 \left( \phi^a (x) \right)^2	+ 
		\sum_{a}\left( f^{man} \right)^2 
			\left( \phi^n (x) - \phi^n_0 
		  \right)^2 + 
	\right.
\nonumber \\
   &&\left.
		\sum_{n}\left( f^{mn8} \right)^2 \left( \phi^8 (x) - \phi^8_0 
		  \right)^2
	\right]
\label{sec4:30}
\end{eqnarray}
which describe the variance of nonlinear oscillations of the gauge field 
and they 
are nonzero inside of the bubble only. It means that the quantized field 
is concentrated in this region. By such a manner the correlation between 
$A^B(x)$ and $A^C(x), B \neq C$ components are nonzero in the same region. 
Consequently one can say that in this approach the quantized field SU(3) 
fields is concentrated in the bubble and can be interpreted as glueball 
in the vacuum. 
\par 
The presented here approach to the QCD is similar to a field correlator 
method \cite{digiacomo} with one difference: in our approach there is 
dynamical equations for the Green's functions which are derived from the 
SU(3) Lagrangian. 
\par 
This approach to the Green's functions which can be approximately considered 
as scalar fields (or a condensate) may have interesting applications for gravity where 
scalar fields have various applications: inflation, boson stars, 
non-Abelian black holes and so on. Our approach allows us to speculate that 
the nonperturbative quantum effects can be very important in some gravitational 
phenomenon. 

\section{Acknowledgments}

This work is supported by ISTC grant KR-677.

\appendix
\section{The effective Lagrangian}
\label{app1}

In order to derive equations describing the quantized field we average 
the Lagrangian over a quantum state $\left.\left. \right| Q \right\rangle$ 
\begin{equation}
\begin{split}
	\left\langle Q \left| \widehat \mathcal{L} \right| Q \right\rangle =&
	\left\langle \widehat \mathcal{L} \right\rangle = 
	\frac{1}{2}	
	\left\langle 
	  \left( \partial_\mu \widehat A^B_\nu  \right) 
	  \left( \partial^\mu \widehat A^{B\nu}  \right) - 
	  \left( \partial_\mu \widehat A^B_\nu  \right) 
	  \left( \partial^\nu \widehat A^{B\mu}  \right) 
	\right\rangle + \\
	&\frac{1}{2}	g f^{BCD} 
	\left\langle 
	  \left( \partial_\mu \widehat A^B_\nu - 
	  \partial_\nu \widehat A^B_\mu \right)
	  \widehat A^{C \mu} \widehat A^{D \nu} 
	\right\rangle + 
	\frac{1}{4}g^2 f^{BC_1D_1} f^{BC_2D_2}
	\left\langle 
	  \widehat A^{C_1}_\mu \widehat A^{D_1}_\nu 
	  \widehat A^{C_2 \mu} \widehat A^{D_2} \nu 
	\right\rangle
\end{split}	
\label{app1:20}
\end{equation}
Schematically we have the following 2, 3 and 4-points 
Green's functions: 
$\left\langle \left( \partial A \right)^2 \right\rangle$, 
$\left\langle \left( \partial A \right) A^2 \right\rangle$ and 
$\left\langle \left( A \right)^4\right\rangle$. At first we introduce  
the 2-point Green's function 
\begin{equation}
	\left\langle 
	  \widehat A^B_\alpha (x) \widehat A^C_\beta (y) 
	\right\rangle = 
	\mathcal{G}^{BC}_{\alpha \beta} (x,y) 
\label{app1:35}	
\end{equation}
The first term on the rhs of equation \eqref{app1:20} is 
\begin{equation}
	\left( \partial_\mu \widehat A^B_\nu (x) \right) 
	\left( \partial^\mu \widehat A^{B\nu}(x) \right) = 
	\partial_{x\mu} \partial_y^\mu 
	\left( \widehat A^B_\nu (x) \right) 
	\left( \widehat A^{B\nu}(y) \right) \Bigr |_{y \rightarrow x} = 
	\eta^{\alpha \beta} \partial_{x\mu} \partial_y^\mu 
	\mathcal{G}^{BB}_{\alpha \beta} (x,y) \Bigr |_{y \rightarrow x} .
\label{app1:60}	
\end{equation}
For the simplicity we consider the case with $x_0=y_0$. For this 
Green's function we use so called one-function approximation 
\cite{vdsin2} 
\begin{equation}
	\mathcal{G}^{AB}_{\alpha \beta} (x,y) \approx 
	-\eta_{\alpha \beta} f^{ACD} f^{BCE} \phi^D (x) \phi^E(y)
\label{app1:80}	
\end{equation}
where $\phi^A(x)$ is the scalar field which describes the 2-point Green's 
function. Physically this approximation means that quantum properties 
of the field $A^B_\mu$ can be approximately described by a scalar field 
$\phi^B(x)$, i.e. in this approximation the Lorentz index $\mu$ is not 
very important. Taking into account this approximation we have 
\begin{equation}
  \left\langle 
	\left( \partial_\mu \widehat A^B_\nu  \right) 
	\left( \partial^\mu \widehat A^{B\nu} \right) 
	\right\rangle = 
	- \eta^\nu_\nu f^{BAC} f^{BAD}
	\left( \partial_\mu \phi^C \right)
	\left( \partial^\mu \phi^D \right) = 
	- 12 \left( \partial_\mu \phi^A \right)
	\left(\partial^\mu \phi^A\right)
\label{app1:90}		
\end{equation}
and 
\begin{equation}
  \left\langle 
	  \left( \partial_\mu \widehat A^B_\nu  \right) 
	  \left( \partial^\nu \widehat A^{B\mu}  \right) 
	\right\rangle =  
	- 3 \left( \partial_\mu \phi^A \right) 
	\left(\partial_\mu \phi^A \right) ,
\label{app1:100}
\end{equation}
Later we suppose that the odd Green's functions can be expressed as the 
sum of the following products 
\begin{equation}
	\left\langle 
	  \widehat A^B_\alpha (x)\widehat A^C_\beta (y)\widehat A^D_\gamma (z)
	\right\rangle \approx 
	\left\langle 
	  \widehat A^B_\alpha (x)\widehat A^C_\beta (y)
	\right\rangle
	\left\langle 
	  \widehat A^D_\gamma (z)
	\right\rangle + \text{(other permutations)}
	= 0 
\label{app1:105}
\end{equation}
as $\langle \widehat A^B_\alpha (x) \rangle = 0$. It gives us 
\begin{equation}
	\left\langle 
	  \left( \partial_\mu \widehat A^B_\alpha (x) \right)
	  \widehat A^C_\beta (x) \widehat A^D_\gamma (x) 
	\right\rangle = 
	\partial_{x\mu} 
	\left\langle 
	  \widehat A^B_\alpha (x) \widehat A^C_\beta (y) \widehat A^D_\gamma (z) 
	\right\rangle \Bigr |_{y,z \rightarrow x} = 0
\label{app1:40}	
\end{equation}
and consequently in our approximation 
\begin{equation}
	\left\langle 
	  \left( \partial_\mu \widehat A^B_\nu - 
	  \partial_\nu \widehat A^B_\mu \right)
	  \widehat A^{C \mu} \widehat A^{D \nu} 
	\right\rangle = 0 .
\label{app1:50}	
\end{equation}
For the last quartic term on the rhs of equation \eqref{app1:20} 
we assume the following approximation 
\begin{equation}
\begin{split}
	&\left\langle 
	  \widehat A^B_\alpha (x) \widehat A^C_\beta (y)
	  \widehat A^D_\gamma (z) \widehat A^R_\delta (u)
	\right\rangle \approx 
	\left\langle 
	  \widehat A^B_\alpha (x) \widehat A^C_\beta (y)
	\right\rangle  
	\left\langle 
	  \widehat A^D_\gamma (z) \widehat A^R_\delta (u)
	\right\rangle + \\
	&\left\langle 
	  \widehat A^B_\alpha (x) \widehat A^D_\gamma (z)
	\right\rangle  
	\left\langle 
	  \widehat A^C_\beta (y) \widehat A^R_\delta (u)
	\right\rangle + 	
	\left\langle 
	  \widehat A^B_\alpha (x) \widehat A^R_\gamma (u)
	\right\rangle  
	\left\langle 
	  \widehat A^C_\beta (y) \widehat A^D_\gamma (z)
	\right\rangle .
\end{split}	
\label{app1:115}
\end{equation}
In fact it is the assumption that 4-point Greens function is the product 
of two 2-points Green's function. In this approximation the lhs of 
\eqref{app1:115} is 
\begin{equation}
\begin{split}
	&\left\langle 
	  \widehat A^B_\mu (x) \widehat A^C_\nu (x)
	  \widehat A^{D\mu} (x) \widehat A^{R\nu} (u)
	\right\rangle = \\
  &\lambda_{1,2;(P_{1,2},Q_{1,2})} 
	\left( 
	  f^{BE_1P_1} f^{CE_1 Q_1} \phi^{P_1}(x) \phi^{Q_1}(x)
	\right)
	\left( 
	  f^{DE_2P_2} f^{RE_2 Q_2} \phi^{P_2}(x) \phi^{Q_2}(x)
	\right) \eta_{\mu\nu} \eta^{\mu\nu} + \\
	&\lambda_{1,2;(P_{1,2},Q_{1,2})} 
	\left( 
	  f^{BE_1P_1} f^{DE_1 Q_1} \phi^{P_1}(x) \phi^{Q_1}(x)
	\right)
	\left( 
	  f^{CE_2P_2} f^{RE_2 Q_2} \phi^{P_2}(x) \phi^{Q_2}(x)
	\right) \eta^\mu_\mu \eta^\nu_\nu + \\
	&\lambda_{1,2;(P_{1,2},Q_{1,2})} 
	\left( 
	  f^{BE_1P_1} f^{RE_1 Q_1} \phi^{P_1}(x) \phi^{Q_1}(x)
	\right)
	\left( 
	  f^{CE_2P_2} f^{DE_2 Q_2} \phi^{P_2}(x) \phi^{Q_2}(x)
	\right) \eta^\nu_\mu \eta^\mu_\nu
\end{split}	
\label{app1:120}
\end{equation}
where $\lambda_{1,2;(P_{1,2},Q_{1,2})}$ is some parameter depending on the values of 
the indices $P_{1,2},Q_{1,2}$ 
\begin{equation}
	\lambda_{1,2; \left (P_{1,2},Q_{1,2} \right )}= 
	\begin{cases}
	  \lambda_1, & \text{if all indices } P_{1,2},Q_{1,2} = 1,2,3, \\
	  \lambda_2, & \text{if all indices } P_{1,2},Q_{1,2} = 4,5,6,7,8, \\
	  1,         & \text{otherwise} 
	\end{cases}
\label{app1:130}
\end{equation}
where $\lambda_{1,2}$ are some parameters. Introducing this index 
$\lambda_{1,2; \left (P_{1,2},Q_{1,2} \right )}$ we would like to say that 
the presented approximate quantization procedure is a little different 
for the scalar field components belonging to the small subgroup 
$SU(2) \in SU(3)$ and the coset $SU(3)/SU(2)$. 
In this case 
\begin{equation}
\begin{split}
	\left( 
	  f^{BE_1P_1} f^{CE_1 Q_1} \phi^{P_1} \phi^{Q_1}
	\right)
	\left( 
	  f^{DE_2P_2} f^{RE_2 Q_2} \phi^{P_2} \phi^{Q_2}
	\right) &= \\ 
	\lambda_1 
	\left( 
	  f^{BE_1a} f^{CE_1 b} \phi^{a} \phi^{b}
	\right)
	\left( 
	  f^{DE_2c} f^{RE_2 d} \phi^{c} \phi^{d}
	\right) + \lambda_2 
	\left( 
	  f^{BE_1m} f^{CE_1 n} \phi^{m} \phi^{n}
	\right)
	\left( 
	  f^{DE_2p} f^{RE_2 q} \phi^{p} \phi^{q}
	\right) &+ \\	
	 (\text{other terms} )&
\end{split}
\label{app1:140}
\end{equation}
The calculations show that 
\begin{eqnarray}
	f^{ABC} f^{ADR}
	\left( 
	  f^{BE_1P_1} f^{CE_1 Q_1} \phi^{P_1} \phi^{Q_1}
	\right)
	\left( 
	  f^{DE_2P_2} f^{RE_2 Q_2} \phi^{P_2} \phi^{Q_2}
	\right) & = & 0, 
\label{app1:150}\\
  f^{ABC} f^{ADR}
	\left( 
	  f^{BE_1a} f^{CE_1 b} \phi^{a} \phi^{b}
	\right)
	\left( 
	  f^{DE_2c} f^{RE_2 d} \phi^{c} \phi^{d}
	\right) & = & 0, 
\label{app1:160}\\
 f^{ABC} f^{ADR}
	\left( 
	  f^{BE_1m} f^{CE_1 n} \phi^{m} \phi^{n}
	\right)
	\left( 
	  f^{DE_2p} f^{RE_2 q} \phi^{p} \phi^{q}
	\right) & = & 0. 
\label{app1:170}
\end{eqnarray}
Consequently 
\begin{equation}
	\lambda_{1,2; \left( P_{1,2},Q_{1,2} \right)} 
	\left( 
	  f^{BE_1P_1} f^{CE_1 Q_1} \phi^{P_1}(x) \phi^{Q_1}(x)
	\right)
	\left( 
	  f^{DE_2P_2} f^{RE_2 Q_2} \phi^{P_2}(x) \phi^{Q_2}(x)
	\right) \eta_{\mu\nu} \eta^{\mu\nu} = 0.
\label{se1:180}
\end{equation}
The similar calculations show that 
\begin{eqnarray}
	f^{ABC} f^{ADR}
	\left( 
	  f^{BE_1P_1} f^{DE_1 Q_1} \phi^{P_1}(x) \phi^{Q_1}(x)
	\right)
	\left( 
	  f^{CE_2P_2} f^{RE_2 Q_2} \phi^{P_2}(x) \phi^{Q_2}(x)
	\right)  & = & 
\nonumber \\
	\frac{27}{8}\left(
	  \phi^a \phi^a + \phi^m \phi^m
	\right)^2 , &&
\label{app1:190}\\
  f^{ABC} f^{ADR}
	\left( 
	  f^{BE_1 a} f^{DE_1 b} \phi^a \phi^b
	\right)
	\left( 
	  f^{CE_2 c} f^{RE_2 d} \phi^c \phi^d
	\right)  & = & 
	\frac{27}{8}\left(
	  \phi^a \phi^a 
	\right)^2, 
\label{app1:200}\\
 f^{ABC} f^{ADR}
	\left( 
	  f^{BE_1 m} f^{DE_1 n} \phi^m \phi^n
	\right)
	\left( 
	  f^{CE_2 p} f^{RE_2 q} \phi^p \phi^q
	\right)  & = & 
	\frac{27}{8}\left(
	  \phi^m \phi^m 
	\right)^2. 
\label{app1:210}
\end{eqnarray}
Consequently 
\begin{equation}
\begin{split}
	\lambda_{1,2; \left( P_{1,2},Q_{1,2} \right)} 
	\left( 
	  f^{BE_1P_1} f^{DE_1 Q_1} \phi^{P_1}(x) \phi^{Q_1}(x)
	\right)
	\left( 
	  f^{CE_2P_2} f^{RE_2 Q_2} \phi^{P_2}(x) \phi^{Q_2}(x)
	\right) &= \\
  \frac{27}{8}\lambda_1 \left(
	  \phi^a \phi^a 
	\right)^2 + 
	\frac{27}{8}\lambda_2 	
	\left(
	  \phi^m \phi^m 
	\right)^2 + 
	\frac{27}{4} \left(
	  \phi^a \phi^a 
	\right)
	\left(
	  \phi^m \phi^m 
	\right) . &
\end{split}
\label{app1:220}
\end{equation}
Analogously
\begin{equation}
\begin{split}
	\lambda_{1,2; \left( P_{1,2},Q_{1,2} \right)} 
	\left( 
	  f^{BE_1P_1} f^{RE_1 Q_1} \phi^{P_1}(x) \phi^{Q_1}(x)
	\right)
	\left( 
	  f^{CE_2P_2} f^{DE_2 Q_2} \phi^{P_2}(x) \phi^{Q_2}(x)
	\right) &= \\
  \frac{27}{8}\lambda_1 \left(
	  \phi^a \phi^a 
	\right)^2 + 
	\frac{27}{8}\lambda_2 	
	\left(
	  \phi^m \phi^m 
	\right)^2 + 
	\frac{27}{4} \left(
	  \phi^a \phi^a 
	\right)
	\left(
	  \phi^m \phi^m 
	\right) . &
\end{split}
\label{app1:225}
\end{equation}
Finally for $x=y=z=u$ the quartic term is 
\begin{equation}
  f^{ARB} f^{ACD}
	\left\langle 
	  \widehat A^{R}_\mu \widehat A^{B}_\nu 
	  \widehat A^{C \mu} \widehat A^{D} \nu 
	\right\rangle = 
	\frac{81}{2} \lambda_1 \left( \phi^a \phi^a \right)^2 + 
	\frac{81}{2} \lambda_2 \left( \phi^m \phi^m \right)^2 + 
	81 \left( \phi^a \phi^a \right) \left( \phi^m \phi^m \right) . 
\label{app1:230}
\end{equation}
Therefore we have the following effective Lagrangian describing 2 and 4-points 
Green's functions 
\begin{equation}
  \mathcal{L}_{eff} = - \frac{9}{2} 
  \left( \partial_\mu \phi^A \right) \left( \partial^\mu \phi^A \right) + 
  \frac{g^2}{4}\left[
    \frac{81}{2} \lambda_1 \left( \phi^a \phi^a \right)^2 + 
	  \frac{81}{2} \lambda_2 \left( \phi^m \phi^m \right)^2 + 
	  81 \left( \phi^a \phi^a \right) \left( \phi^m \phi^m \right) 
	\right].
\label{app1:240}	
\end{equation}
If we redefine $\phi^a \rightarrow 2 \phi^a /(3 g)$ and 
$\lambda_{1,2} \rightarrow \lambda_{1,2} /2$ we will have the ordinary Lagrangian 
for the scalar field 
\begin{equation}
	\frac{g^2}{4} \mathcal {L}_{eff} = - \frac{1}{2}
	\left( \partial_\mu \phi^A \right) \left( \partial^\mu \phi^A \right)+ 
	\frac{\lambda_1}{4} \left( \phi^a \phi^a \right)^2 + 
	\frac{\lambda_2}{4} \left( \phi^m \phi^m \right)^2 + 
	\left( \phi^a \phi^a \right) \left( \phi^m \phi^m \right) .
\label{app1:250}	
\end{equation}
Now it is necessary to do an essential remark. The SU(3) Lagrangian 
for the gauge group $A^B_\mu$ is very nonlinear: it has $A^4$ terms.
It is well known \cite{coleman1} that in $\lambda \phi^4$ theory 
the similar nonlinearity 
give rise to an additional term to potential term. One can suppose 
that the similar situation takes place in this situation, too. Here we 
suppose that the nonlinear terms like $A^4$ leads to the appearance 
of some term in the initial Lagrangian. For the simplicity we assume 
that the mass term will appear. Thus the final form of the effective 
Lagrangian is
\begin{equation}
\begin{split}
	&\frac{g^2}{4} \mathcal {L}_{eff} = - \frac{1}{2}\left( \partial_\mu \phi^A \right)^2 + 
	\frac{\lambda_1}{4} 
	\left[ \phi^a \phi^a - \phi^a_0 \phi^a_0 
	\right]^2 - \frac{\lambda_1}{4} \left( \phi^a_0 \phi^a_0  \right)^2 + \\
	&\frac{\lambda_2}{4} 
	\left[ \phi^m \phi^m - \phi^m_0 \phi^m_0 
	\right]^2 - \frac{\lambda_2}{4} \left( \phi^m_0 \phi^m_0  \right)^2 + 
	\left( \phi^a \phi^a \right) \left( \phi^m \phi^m \right) 
\end{split}
\label{app1:260}
\end{equation}
where $\phi^A_0$ are some constants. In this situation the field equations 
for the approximate scalar description of the QCD are  
\begin{eqnarray}
  \partial_\mu \partial^\mu \phi^a &=& 
  \phi^a \left[ 2 \phi^m \phi^m + \lambda_1 
  \left(
    \phi^a \phi^a- \phi^a_0 \phi^a_0 
  \right) \right],
\label{app1:270}\\
  \partial_\mu \partial^\mu \phi^m &=& 
  \phi^m \left[ 2 \phi^a \phi^a + \lambda_2 
  \left(
    \phi^m \phi^m- \phi^m_0 \phi^m_0 
  \right) \right].
\label{app1:280}
\end{eqnarray}
In conclusion we have to note that this procedure for the approximate 
calculations of 2 and 4-points Green's functions should be some 
approximation for an exact procedure which obtains \textit{all} 
Green's functions bu a nonperturbative manner. At first such procedure 
was offered by Heisenberg for the quantization of a nonlinear spinor field 
\cite{heis} and later was applied for the QCD \cite{vdsin2}.

\section{The numerical calculations of the soliton}
\label{app2}

For the validation of the presented method of solving the nonlinear equations 
\eqref{sec2:30} \eqref{sec2:40} we choose a soliton solution. The corresponding 
equation is 
\begin{equation}
	\frac{d^2 y}{d x^2} = y'' = y \left( 1 - y^2 \right).
\label{app2:10}
\end{equation}
The solution is 
\begin{equation}
	y(x) = \frac{\sqrt{2}}{\cosh x}.
\label{app2:20}
\end{equation}
We rewrite the \eqref{app2:10} equation in the form of the Schr\"odinger 
equation 
\begin{equation}
	-y'' + y V_{eff} = - \lambda y
\label{app2:30}
\end{equation}
where $V_{eff} = -y^2$ and $\lambda = 1$. It shows us that the regular 
solution exists only for a discrete spectrum of ``energy level'' 
$\lambda$. We will solve this equation by an iterative procedure. 
At first we have the equation 
\begin{equation}
	 -y_1'' + y_1 \left( -y_0^2 \right) = 
	 - \lambda_1 y_1 
\label{app2:50}
\end{equation}
for the first approximation $y_1(x)$ and where $\lambda_1$ is the first 
approximation for the $\lambda$. For the numerical solution we choose 
the null approximation as 
\begin{equation}
	y_0 =  \frac{\sqrt{2}}{\cosh \left( \frac{x}{2} \right)}.
\label{app2:40}
\end{equation}
The typical solution for the arbitrary values of the parameter 
$\lambda_1$ is presented on Fig.\ref{fig:soliton-sing}. 
\begin{figure}[h]
  \begin{minipage}[t]{.45\linewidth}
  \begin{center}
    \fbox{
    \includegraphics[height=5cm,width=5cm]{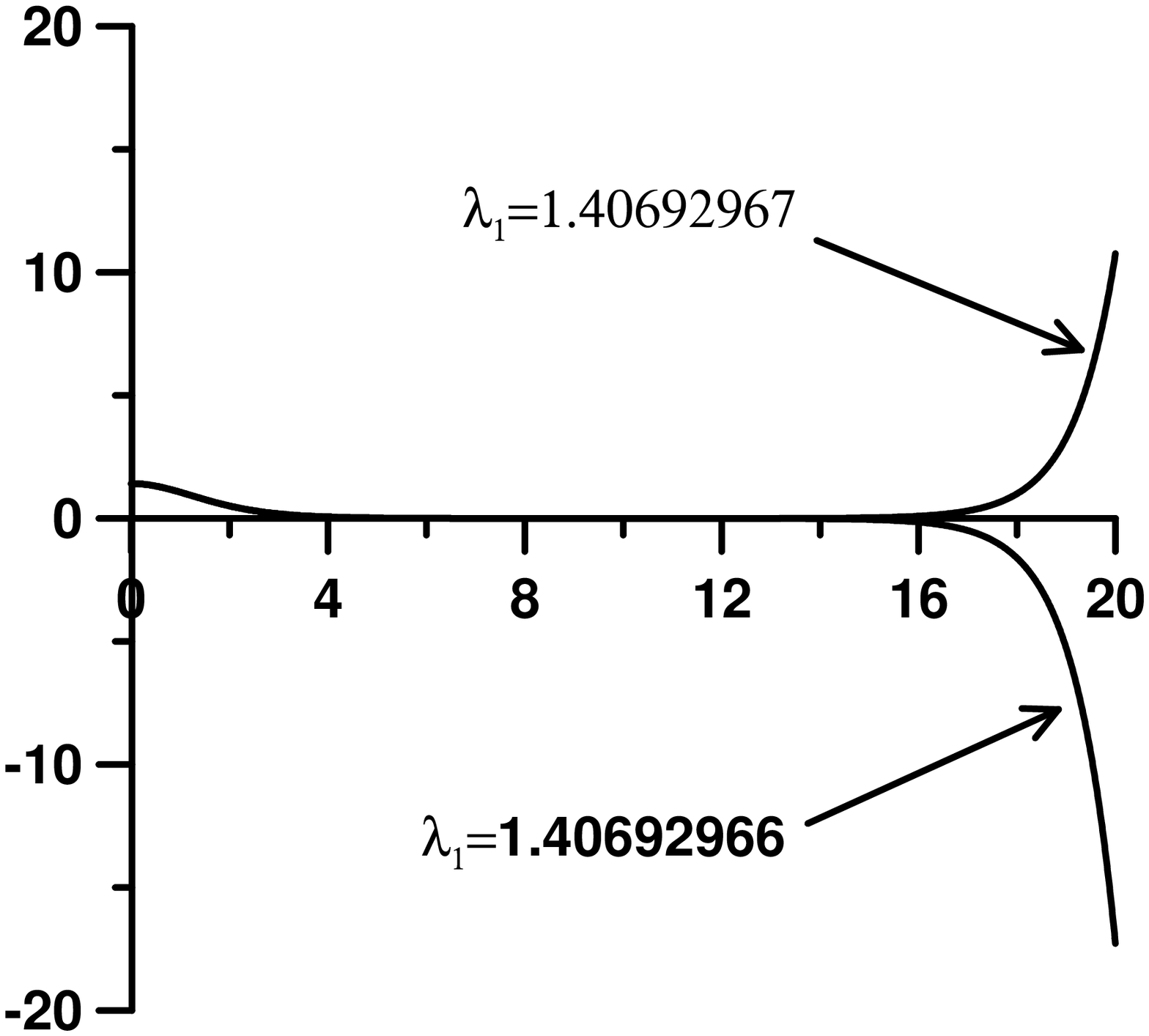}}
    \caption{The singular solutions for the soliton equation.}
    \label{fig:soliton-sing}
  \end{center}  
  \end{minipage}\hfill
  \begin{minipage}[t]{.45\linewidth}
  \begin{center}
    \fbox{
    \includegraphics[height=5cm,width=5cm]{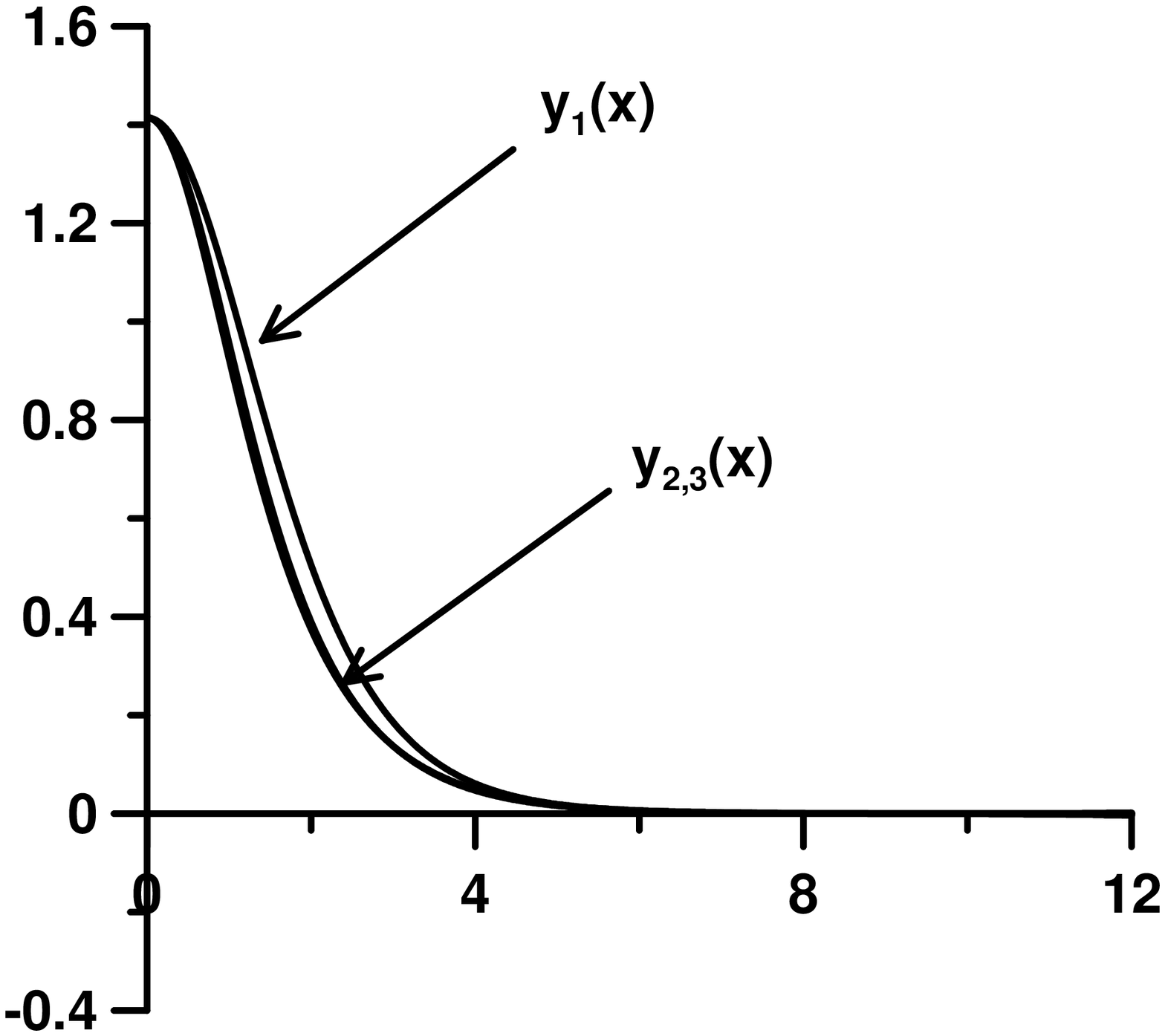}}
    \caption{The iterative functions $y_{1,2,3}$.}
    \label{fig:soliton-reg}
  \end{center}  
  \end{minipage} 
\end{figure}
This picture shows us that 
there is a value $\lambda^*_1$ for which the solution is exceptional one. 
One can find this exceptional solution choosing the appropriate value 
of the ``energy level'' $\lambda_1^*$. After which an exceptional solution 
$y^*_1(x)$ is substituted into equation for the second approximation 
$y_2(x)$
\begin{equation}
	 -y_2'' - y_2 \left( y^*_1 \right)^2 = - \lambda_2 y_2
\label{app2:60}
\end{equation}
and so on. The result is presented on Table \ref{table2} and 
Fig.\ref{fig:soliton-reg}. One can see that $\lambda^*_i \rightarrow 1$ 
and $y^*_i(x)$ is convergent to $y^*(x)$. 
\begin{table}[h]
    \begin{center}
        \begin{tabular}{|c|c|c|c|}\hline
          i & 1 & 2& 3 \\ \hline
            $\lambda^*_i$ & 1.406929666\ldots & 1.148564915\ldots & 1.03968278\ldots 
            \\ \hline
        \end{tabular}
    \end{center}
    \caption{The iterative values of the parameter $\lambda$.}
    \label{table2}
\end{table}

\end{document}